\documentclass[runningheads]{llncs}
\usepackage[hidelinks]{hyperref}
\usepackage{lineno,}
\usepackage{enumitem}
\usepackage{amsmath}
\usepackage{amsfonts}
\usepackage{wrapfig}
\usepackage{color}

\modulolinenumbers[5]
\usepackage{tabu}
\usepackage{adjustbox}
\usepackage{color, colortbl}
\definecolor{Gray}{gray}{0.9}
\definecolor{White}{rgb}{1,1,1}
\definecolor{LightCyan}{gray}{0.9}
\usepackage[first=0,last=9]{lcg}

\usepackage[skip=0.5\baselineskip]{caption}
\usepackage{graphicx}
\usepackage{pdflscape}
\begin{document}
\title{Large-Scale Shill Bidder Detection in E-commerce}
%
%
\author{Michael Fire\inst{1} \and
Rami Puzis\inst{1}\and
Dima Kagan\inst{1} \and
Yuval Elovici\inst{1}}

\authorrunning{Fire et al.}
%
\institute{Ben-Gurion University of the Negev\\
\email{\{mickyfi,puzis,kagandi,elovici\}@post.bgu.ac.il}}
\maketitle              
\begin{abstract}
User feedback is one of the most effective methods to build and maintain trust in electronic commerce platforms. 
Unfortunately, dishonest sellers often bend over backwards to manipulate users' feedback or place phony bids in order to increase their own sales and harm competitors. 
The black market of user feedback, supported by a plethora of shill bidders, prospers on top of legitimate electronic commerce. 
In this paper we investigate the ecosystem of shill bidders based on large-scale data by analyzing hundreds of millions of users who performed billions of transactions, and we propose a machine-learning-based method for identifying communities of users that methodically provide dishonest feedback.
Our results show that (1) shill bidders can be identified with high precision based on their transaction and feedback statistics; and (2) in contrast to legitimate buyers and sellers, shill bidders form cliques to support each other.

\end{abstract}

\section{Introduction}
\label{sec:introduction}
Electronic commerce (e-commerce) usage has increased sharply as e-commerce platforms have become interwoven into people's everyday lives as places to buy and sell products.
The e-commerce worldwide sales to consumers is expected to pass the four trillion dollars mark in 2020, almost quadrupling itself compared to 2014~\cite{Globalr44:online,GlobalEc74:online}. E-commerce platforms, such as Amazon,\footnote{\url{http://www.amazon.com/}} Alibaba,\footnote{\url{http://www.alibaba.com/}} eBay,\footnote{\url{http://www.ebay.com/}} Etsy,\footnote{\url{http://www.etsy.com/}} and OnlineAuction,\footnote{\url{http://www.onlineauction.com/}} already have hundreds of millions of active users~\cite{alibabausers2012,ebayusers2013,amazonusers2013,etsy2012}.
In many e-commerce platforms, users can use the platforms to sell and buy various products from each other, from simple everyday products, such as a cup holder which costs only a few dollars, to more exotic products, such as a fighter jet which costs several million dollars~\cite{jet2004}. 
Furthermore, more than 25 million individuals use e-commerce platforms as their primary or secondary source of income \cite{ThereAre73:online}.

In many e-commerce platforms, after a user purchases a product, he or she can rate the product and write a review regarding the product's features, such as the product's quality~\cite{mudambi2010makes}. Furthermore, many e-commerce platforms employ reputation systems, in which the buyer can leave feedback regarding the seller of the product~\cite{resnick2000reputation}. 
The accumulated feedback on each seller, which in many cases is viewable to other users of the website, can assist other users in building trust towards the seller~\cite{resnick2000reputation}. 
According to Lucking-Reile et al.~\cite{lucking2007pennies} the feedback these sellers receive can have a measurable effect on their auction prices, where negative feedback has a much greater effect than positive feedback ratings.
Moreover, in cases where sellers receive too much negative feedback from other users, the website operator can decide to revoke the seller's selling privileges.
For example, Facebook banned and reduced the number of ads for businesses who received too much negative feedback from buyers~\cite{Facebook24:online}.

As in many other online platforms, such as search engines, online social networks, and online gaming platforms, the platform's users can utilize dishonest techniques to manipulate the platform's statistics in order to create profits~\cite{Kabus:2005:ACD:1103599.1103607}.
Moreover, in many online platforms, the platform's users can purchase this type of service from third-party providers, which in most cases violates the platforms terms of service. For example, the market of buying fake followers and fake retweets in Twitter is already a multimillion-dollar business~\cite{twitterny}. 
The equivalent to Twitter's fake followers in e-commerce are fake reviews.
The fake review industry is flourishing; there are many paid reviewers and even people who receive free products in exchange for a positive review \cite{AmazonsF7:online,IGetPaid40:online}.
In e-commerce platforms dishonest users (referred to as shill bidders) can increase their product's price in auctions by bidding on products with the intent to artificially increase the product's price or desirability~\cite{ebay_shillpolicy,trevathan2009detecting}. 
Additionally, shill bidders are also users who buy products in order to artificially improve a seller's feedback or the product's search standing~\cite{ebay_shillpolicy}.
Shill bidding is forbidden in many e-commerce platforms, such as eBay~\cite{ebay_shillpolicy}, tophatter \cite{Tophatte74:online}, flippa \cite{WhatisSh31:online}, etc. Moreover, shill bidding in online auctions is illegal in some countries, such as the United States, and can be considered as wire fraud, a felony which can lead to a maximum penalty of up to four years in prison and a million dollars in fines~\cite{ny2001shill}.  

\begin{figure}[t]
	\begin{center}
		\includegraphics[width=0.5\textwidth]{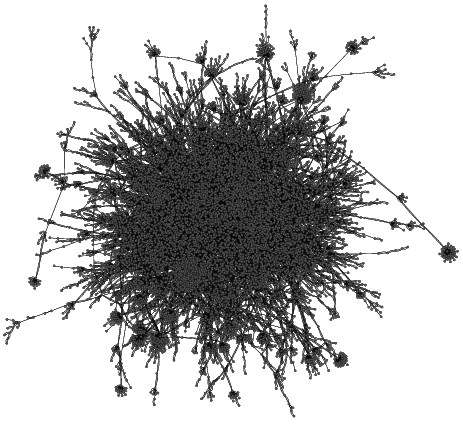}
	\end{center}
	\caption{ \label{fig:graph}
		Shill bidders feedback weighted directed graph (subgraph with 10,249 vertices and 21,859 links.)
	}
\end{figure}

In this paper we study the shill bidder ecosystem. 
The main contributions of this study are threefold. 
First, we offer generic algorithms for identifying shill bidders in e-commerce platforms. 
Moreover, we evaluate these algorithms on one the biggest e-commerce datasets in the world (referred to as the \textit{e-commerce dataset}), which includes several billion buying transactions and several billion feedback interactions between the platform's users. 
Second, we analyze the activities of over 187,244 identified shill bidders to better understand their characteristics (see Figure~\ref{fig:graph}). 
Lastly, we investigate the ecosystem of shill bidders and offer methods on how this ecosystem's properties can be utilized to better identify shill bidders. To the best of our knowledge, this study which aims to identify shill bidders is the largest of its kind to date.

The remainder of the paper is structured as follows: In Section~\ref{sec:related} we give a brief overview of previous relevant studies related to e-commerce security and shill bidders.
In this section, we also introduce several studies which used similar
data mining algorithms as this study in other online platforms, such as Facebook and Twitter. Next, in Section~\ref{sec:methods} we present the methods and
algorithms we developed for identifying shill bidders.
Afterwards, in Section~\ref{sec:empirical} we describe the e-commerce dataset.
Additionally, in Section~\ref{sec:empirical}, we present the performed empirical evaluation process and the results of the shill bidder identification algorithms on the e-commerce dataset.
Then, in Section~\ref{sec:eco}, we present the performed empirical study of the shill bidders' characteristics.
Furthermore, in this section we also present our analysis of the shill bidders' ecosystem.
Lastly, in Section~\ref{sec:conclusion}, we present our conclusions and also offer future research directions.

\section{Related Work}
\label{sec:related}
\subsection{E-commerce Platform Fraud }
In the past decade, during the rise in the availability of the Internet, there was also a constant growth in popularity and volume of e-commerce~\cite{forbesecommerc}. 
However, the extreme popularity and the high anonymity of e-commerce attracted many fraudsters.
According to Javelin's financial impact of fraud study~\cite{Financia31:online}, the 2017 online revenue loss estimation due to fraud stood at 40 billion USD.

In the past several years, many research efforts have been devoted to studying e-commerce platform's security and trust~\cite{Zhang2013299}; however, only a small portion of this work centered on the detection of shill bidders. 
Chakraborty et al.~\cite{chakraborty2004auctions} defined shill bidding as an illicit participation of sellers in auctions designed to increase the price at which a product sells.
There are many types of shill bidding, and we are going to present two notable types as examples:

The first type is competitive shilling, which is the simplest kind of shill.
Kauffman et al.~\cite{kauffman2003running} explains that it is used to make bidders pay more for a product. 
Basically the seller or his or her accomplices enter bids to make legitimate bidders pay more.
The second type is reserve price shilling.
Kauffman et al.~\cite{kauffman2003running} notes that this is a shilling that can be used to avoid paying auction house fees such as insertion fees or secret reserve fees.
In addition, there are questionable techniques that stand on a thin line between tips or tricks and shill bidding. 
For example, Roth and Ockenfels~\cite{roth2000last} describe last-minute bidding (``sniping''), which is basically placing your bid exactly before the auction ends in order to pay a lower price and snatch the product.
There is controversy about sniping and it is considered legal~\cite{makeuseof,techopedia}. 
Moreover, Roth and Ockenfels~\cite{roth2000last} note that late bidding may be used by a dishonest seller who attempts to raise the price by using shill bidders.

The classical approach for tackling the shill-bidding problem is to use detection and prediction techniques.
In 2005, Chau and Faloutsos~\cite{chau2005fraud} purposed a method for fraudster detection in online auctions.
First, they created different combinations of user information-based features, for example, the number of items bought or sold in a certain time period.
The features were extracted from a dataset of 43 fraudsters and 72 legitimate users from eBay.
Next, they inserted the feature sets into a C5.0 decision-tree-based classification algorithm.
In their best combination of features, they were able to detect malicious users with a precision of 82\%, a True Positive (TP) rate of 83\%, and a False Positive (FP) rate of 11\%.
In 2006, Chau et al.~\cite{chau2006detecting} presented an expansion of their first work, a novel 2-Level Fraud Spotting (2LFS) model.
The 2LFS uses two types of features: user information features and network topography features.
Afterwards, they used a Belief Propagation algorithm over a Markov Random Field to create the detection model.
Finally, they evaluated their model on a manually labeled dataset of 55 users (graph with 55 nodes and 620 edges). 
In this work they demonstrated that social network information detection techniques can be applied in e-commerce fraud detection.

In 2007, Pandit~\cite{pandit2007netprobe} developed a model that represents users and transactions as a Markov Random Field.
The model uses a Belief Propagation mechanism in order to detect fraudulent users.
The detected fraudsters with a precision of 0.9 for the synthetic dataset; however, they were unable to present algorithm evaluation results on real-world datasets because they were unable to label them.
In 2008, Beyene et al.~\cite{beyene2008ebay} presented a model that represents eBay transactions and feedback in a graph.
In their study they explained the difference between the eBays graph and a standard social network graph.
First of all, they found that a rich club phenomenon~\cite{zhou2004rich} did not exist in the eBay feedback graph.
Second, they showed that preferential attachment holds only partially.
In addition, they discovered that when a user is sufficiently trustable, the exact number of positive reviews became less important.
However, negative reviews significantly hurt the user's credibility.	

In 2011, Chang et el.~\cite{Chang201111244} purposed a new two-stage phased model for early fraud detection in online auctions.
The model's first phase constructed behavior models based on users' transaction histories.
The first phase's main was feature extraction in a way that significant behavioral differences between legitimate users and fraudsters could be determined.
The second phase was fraud detection where the data of a suspicious account was inserted into the detection model in order to test if the seller is legitimate or not.
They performed an Instance-Based Learning (IBL) algorithm.
Moreover, the modeling here was based on phased modeling, which means that for each part of the data lifespan a behavior model is built, and also a separate detection model is built for each of the models.
They were able to get an average recall rate of classifying accounts of over 93\%.

In 2014, Tsang et al.~\cite{tsang2014detecting} proposed a detection method based on supervised learning with generated synthetic data.
In their study they proved that using synthetic data can have advantages over previous work using real data.
They tested decision tree and neural network models with different kinds of features on synthetic and commercial datasets, where the largest dataset contained  58,162 users.
They found the best results by using a decision-tree-based model with 0.999 and 0.977 detection precision for simple and complex shill cases, respectively.

In 2016 Majadi et al. \cite{majadi2016analysis} analyzed previous studies on shill bidders and found that the most common features used in past literatures were: first bidding, last bidding, bid increment, outbid time, bid frequency, affinity to the sellers, and winning ratio.

In 2017 Kaghazgaran et al. \cite{kaghazgaran2017behavioral} analyzed fake-review properties.
They found that fake reviewers give longer reviews and they tend write their reviews in bursts.

In 2018, Ganguly et al. \cite{ganguly2018online} proposed an SVM-based method for shill bidder detection.
They evaluated their method on a dataset that contained information on 149 auctions and 1024 bidders who bid on PDAs in eBay.
In order to label the dataset, they used hierarchical clustering and manually labeled clusters that looked suspicious. 
Their classifier achieved AUC an of 0.86 using 10-fold cross validation. 

\subsection{Identifying Malicious Users in Online Platforms Using Supervised Learning}
In addition to e-commerce, there are several other platforms that have risen during the internet era.
These platforms have enormous numbers of users.
For example, Facebook, the biggest online social network, has more than 2.23 billion monthly active users~\cite{facebook-newsroom} and the user number is still growing.
The similarity between these platforms is not only their size and popularity, but also the high resemblance in the nature of their malicious users. The e-commerce malicious users are very similar to spammers and fake profiles in social networks, as well as bots in online games.
These malicious users have a goal that is different from benign users, and they perform non-standard actions with some hidden motives.
A report~\cite{Facebook18:online} suggests that the current number of fake or clone accounts on Facebook is between 13\% and 14\%.
Until recently, Twitter was considered a controversial social network that was a safe haven for bots and fake profiles \cite{WhyTwitt48:online}.
In the past year, Twitter started a full-scale campaign against fake users and bots and deleted about 6\% of its users \cite{Battling37:online,Exclusiv23:online}.

In 2012, Rahman et al.~\cite{rahman2012frappe} developed the FRAppE application for socware identification.
They used various application properties, such as the number of permissions required by the applications, as classifier features. 
Rahman et al. used Support Vector Machine (SVM) to build a classifier for socware detection. 
They detected socware with 99.5\% accuracy and a low FP rate of 4.1\%.
In 2010, Wong~\cite{wang2010don} presented a method for spammer identification on Twitter.
His method was based on the use of Twitter graph structure in order to build a classifier that was based on graph features.
This method was proven to be fairly successful with 89\% precision.
Another interesting study was performed by Mitterhofer et al.~\cite{mitterhofer2009server}.
They developed a method of bot detection in World of Warcraft.
Their detection mechanism was based on the route that the character made in the game.
They discovered that bots did repetitive actions like traveling the same route many more times than a benign user.

The general methodology of using supervised learning algorithms for classifying users in online platforms covers the following topics: 
feature extraction, ground truth classification, choice of a learning algorithm, choice of a training set, evaluation method, and performance metrics. 
When the data is imbalanced, e.g., malicious users are underrepresented in the data, a training set that reflects the real distribution of users may result in poorly trained classifiers. 
Tsang et al.~\cite{tsang2014detecting} suggested that by changing the ratios of legitimate and malicious users we can improve the TP rate and the FP rate.
They also mentioned that the training set size can outweigh the effects of the class imbalance.
Chawla et al.~\cite{chawla2004editorial} noted that measures like accuracy can be misleading, and they suggested using more accurate measures such as Receiver Operating Characteristic (ROC) curves.
Another method that Chewla et al. suggested is to perform oversampling or undersampling in order to minimize the imbalance effect on the classification result.
The main idea in oversampling/undersampling is to fit class distribution of the dataset to the class in the real world.
In 2016 Hooi et al. \cite{hooi2016fraudar} developed FRAUDAR, a system for identifying “camouflaged” malicious accounts. FRAUDAR uses density-based features in order to identify malicious vertices in bipartite networks.

In 2017 Kumar et al. \cite{kumar2017army} studied sockpuppetry in discussion communities.
They discovered that sockpuppets behave differently from benign users; for instance, they have more clustered ego-networks and are more likely to interact with each other. 

\section{Identifying Shill Bidders Using Supervised Learning}
\label{sec:methods}
\label{sec:ml}

In this study, we focus on developing generic classifiers that, can identify shill bidders in various e-commerce platforms. 
To cope with the challenge of identifying shill bidders in e-commerce platforms, we follow the regular methodology of using supervised learning algorithms for predicting the likelihood of a user to be a shill bidder. 
User features employed in this study to identify shill bidders are common to many e-commerce platforms.

\subsection{Feature Extraction}
\label{sec:features}

In order to construct classifiers for identifying shill bidders, we define features to be extracted from the transaction and feedback data of each e-commerce user. 
Next, we describe in detail each one of the features we extract for every e-commerce user.

Let be $v$ an user in an e-commerce platform. 
For each user $v$, we can extract features based on $v$'s buying and selling transactions, such as $v$'s number of buying transactions, and based on $v$'s given and received feedback, such as the amount of feedback $v$ gave. 
In this section, we describe in detail all the features we extracted and used during this study's experiments. We open this section with the formal definitions of the extracted features which were based on the users' buying and selling transactions. Afterwards, we introduce the formal definition of all the extracted features that were based on the users' feedback activities. 
Next, we present several personal user's features, such as the country the user's declared to live in.
Lastly, we present the target class feature.

\textbf{Transaction Features.} Let $G_T = <V, E_T>$ be the directed multigraph that represents the buying and selling transactions between two users in the e-commerce platform, where $V$ is the multigraph vertices set, which contains all the e-commerce users, and $E_T$ is the multigraph's links set, which contains data on all the transaction interactions between e-commerce users.
The links in the transactions multigraph are denoted by $e_T:=(u,v,p,d) \in E_T$, where $u,v \in V$ are two e-commerce users, $p$ is the product which was purchased, and $d$ is the purchase time and date. Each link $e_T$ represents a buying transaction of a single product $p$, which a user $u$ bought from a user $v$ in time and date $d$. For each link $e_T \in E_T$, we also define the following three additional properties:
\begin{enumerate}
 \item $e^q_T$ the transaction's product quantity. Namely, the purchased quantity of the product in the transaction.
 \item $e^p_T$ the purchased product price in US dollars.
 \item $e^a_T := e^q_T \cdot e^p_T$ - the transaction total amount in US dollars.
\end{enumerate}
 Using these definitions, we define the features for each $v \in V$ (see Table \ref{tab:features}).
 
\begin{table*}[t]
\caption{ \normalsize \label{tab:all-features} Features  }
\scriptsize
 \scalebox{0.6}{\begin{tabular} {  m{3.5cm} m{10cm} m{5.5cm} } 
 \hline
 \textbf{Name}& \textbf{Description}  & \textbf{Formula}  \\ \hline 
  \multicolumn{3}{c}{Transaction Features} \\
  \hline 
\rowcolor{LightCyan}
\textbf{Buy-Trans-Num($v$)}& The total number of buying transactions which $v$ performed.
	With respect to $G_T$'s topology, the Buy-Trans-Num($v$) feature is equal to the out-degree of $v$ in the multigraph $G_T$.   & $ \mid \{(v,u,p,d) \in E_T  \mid  \exists u \in V\}  \mid   $  \\ 
 \textbf{Sell-Trans-Num($v$)}  & The total number of selling transactions which $v$ performed. 
	With respect to $G_T$'s topology, the Sell-Trans-Num($v$) feature is equal to the in-degree of $v$ in the multigraph $G_T$. & 	$  \mid \{(v,u,p,d) \in E_T  \mid  \exists u \in V\}  \mid  $ \\   \rowcolor{LightCyan}
  \textbf{Unique-Sellers($v$)}  & The distinct number of users which $v$ bought products from.
	With respect to $G_T$'s topology, the Unique-Sellers($v$) feature is equal to the number of vertices in the multigraph $G_T$ which are connected to $v$ by at least one out-link. & $   \mid \{ u \in V  \mid  \exists (u,v,p,d) \in E_T \}  \mid  $ \\  
  \textbf{Unique-Buyers($v$)} & The distinct number of users which $v$ sold products to.
	With respect to $G_T$'s topology, the Unique-Buyers($v$) feature is equal to the number of vertices in the multigraph $G_T$ which are connected to $v$ by at least one in-link.  & $  \{ u \in V  \mid  \exists (v,u,p,d) \in E_T \}  \mid  $ \\   \rowcolor{LightCyan}
  \textbf{Bidir-Trans-Users($v$)}  & The distinct number of users which $v$ sold products to, and also bought products from.  & $  \mid \{ u \in V  \mid  \exists (u,v,p,d), (v,u,p,d) \in E_T \}  \mid $  \\ 
 \textbf{Max-Buy-Price($v$)}   & The maximal paid amount in USD which $v$ paid to another user in a single buying transaction. & 	$
	 max(\{ e^a_T  \mid  \exists e_T:=(v,u,p,d) \in E_T, u \in V\})  
	$	   \\   \rowcolor{LightCyan}
  \textbf{Min-Buy-Price($v$)}   & The minimal paid amount in USD which $v$ paid to another user in a single buying transaction. & 	  	$
	min(\{ e^a_T  \mid  \exists e_T:=(v,u,p,d) \in E_T, u \in V\})  
	$\\  

 \textbf{Max-Buy-Quantity($v$)}& 	
 $
	 max(\{ e^q_T  \mid  \exists e_T:=(v,u,p,d) \in E_T, u \in V\})
	$	   \\   \rowcolor{LightCyan}
 \textbf{Total-Buy-Quantity($v$)}     & The maximal number of products which $v$ bought from another user in a single buying transaction. & 	$  \sum_{\{ e_T=(v,u,p,d) \in E_T  \mid  u \in V\}}e^q_T  $	   \\ 

  \textbf{Total-Buy-Amount($v$)}    & The total amount in USD which $v$ bought in products from other users. & 	$  \sum_{\{ e_T=(v,u,p,d) \in E_T  \mid  u \in V\}}e^a_T  $
		   \\   \rowcolor{LightCyan}
    \textbf{Max-Sell-Price($v$)}  & The maximal amount in USD which $v$ received from another user in a single selling transaction. & 	$
	 max(\{ e^a_T  \mid  \exists e_T:=(u,v,p,d) \in E_T, u \in V\})
	$	   \\  
     \textbf{Min-Sell-Price($v$)} & The minimal amount in USD which $v$ received from another user in a single selling transaction. & $ min(\{ e^a_T  \mid  \exists e_T:=(u,v,p,d) \in E_T, u \in V\})$	   \\   \rowcolor{LightCyan}
      \textbf{Max-Sell-Quantity($v$)}  & The maximal number of products which $v$ sold to another user in a single buying transaction.& 		$
	 max(\{ e^q_T  \mid  \exists e_T:=(u,v,p,d) \in E_T, u \in V\})$ \\  
      \textbf{Total-Sell-Quantity($v$)} & The overall number of products which $v$ sold to other users. & $\sum_{\{ e_T=(v,u,p,d) \in E_T  \mid  u \in V\}}e^q_T  $	   \\   \rowcolor{LightCyan}
     \textbf{Total-Sell-Amount($v$)}  & The total amount in USD which $v$ sold in products to other users. & $\sum_{\{ e_T=(u,v,p,d) \in E_T  \mid  u \in V\}}e^a_T  $	   \\  \hline
     \multicolumn{3}{c}{Feedback Features} \\

 \hline \rowcolor{LightCyan}

\textbf{Gvn-Fdbk-Num($v$)} & The total amount of feedback a user $v$ gave to other users.  & $ \mid \{(v,u,r,d) \in E_F \mid \forall u \in V \} \mid $ \\  
\textbf{Rcv-Fdbk-Num($v$)} & The total amount of feedback a user $v$ received from other users. &  $ \mid \{(u,v,r,d) \in E_F \mid \forall u \in V \} \mid $ \\  \rowcolor{LightCyan}
\textbf{Gvn-Unique-Fdbk($v$)} & The number of unique users which received feedback from $v$.  & $ \mid \{u \in V \mid \exists (v,u,r,d) \in E_F\} \mid $ \\  
\textbf{Rcv-Unique-Fdbk($v$)}  & The number of unique users, which gave feedback to $v$. & $  \mid \{u \in V \mid \exists (u,v,r,d) \in E_F\} \mid $ \\   \rowcolor{LightCyan}
\textbf{Bidir-Fdbk-Users($v$)}  & The distinct number of users which $v$ gave feedback to, and also received feedback from.  & $ \mid \{ u \in V  \mid  \exists (u,v,p,d), (v,u,p,d) \in E_F \}  \mid $ \\  
\textbf{Gvn-Pos-Fdbk($v$)}  & The number of positive feedback ratings a user $v$ received from other users.   & $ \mid \{(v,u,r,d) \in E_F \mid r \textgreater 0 \} \mid $ \\  \rowcolor{LightCyan}
\textbf{Gvn-Neg-Fdbk($v$)}  & The number of negative feedback ratings a user $v$ received from other users. &  $ \mid \{(v,u,r,d) \in E_F \mid r \textless 0 \} \mid $\\  
\textbf{Rcv-Pos-Fdbk($v$)}  & The number of positive feedback ratings a user $v$ received from other users. & $  \mid \{(u,v,r,d) \in E_F \mid r \textgreater 0 \}  \mid $ \\  \rowcolor{LightCyan}
\textbf{Rcv-Neg-Fdbk($v$)}  & The number of negative feedback ratings a user $v$ received from other users.  &  $ \mid \{(u,v,r,d) \in E_F \mid r \textless 0 \}  \mid $\\  
\textbf{Gvn-Fdbk-RSum($v$)}  & The sum of the feedback ratings a user $v$ gave to other users.& $\sum_{\{(v,u,r,d) \in E_F \mid \forall u \in V \}} r$ \\   \rowcolor{LightCyan}
\textbf{Rcv-Fdbk-RSum($v$)}  & The sum of the feedback ratings a user $v$ received from other users. &  $\sum_{\{(u,v,r,d) \in E_F \mid \forall u \in V \}} r$ \\  
\textbf{Gvn-Fdbk-Avg($v$)}  & The average of the feedback ratings a user $v$ gave to other users.&  $\frac{Gvn-Fdbk-RSum(v)}{Gvn-Fdbk-Num(v)}$ \\  \rowcolor{LightCyan}
\textbf{Rcv-Fdbk-Avg($v$)}  & The average of the feedback ratings a user $v$ received from other users.  &   $\frac{Rcv-Fdbk-RSum(v)}{Rcv-Fdbk-Num(v)}$ \\\hline 

\end{tabular}}
\end{table*}

\textbf{Feedback Features.} Similar to the transactions-directed multigraph, we can also define the feedback-directed multigraph, which is based on the e-commerce users' feedback.
Formally, let $G_F = <V, E_F>$ be the directed multigraph that represents the feedback activities between two users in the e-commerce platforms, where $V$ is the multigraph vertices set, which contains all the e-commerce users, and $E_F$ is the multigraph's links set, which contains data on all the feedback interactions between e-commerce users.
The links in the feedback multigraph are denoted by $e_F:=(u,v,r,d) \in E_F$, where $u,v \in V$ are two e-commerce users, $ r \in \mathbb{Z}$ is the feedback rating, and $d$ is a the feedback's time and date. Each link $e_F$ represents a feedback interaction with rating of $r$, which a user $u$ gave a user $v$ in time and date $d$.\footnote{It is worth to mentioning that some e-commerce platforms provide the sellers the opportunity to also rank the buyers. In this study, we treated feedback which was given from a seller to a buyer as the same as feedback from a buyer to a seller. }
Using these feedback-directed multigraph definitions, we define features for each user $v \in V$ (see Table \ref{tab:features}).

\textbf{Users Details Features.} To construct our supervised learning classifiers, we also utilized the following users' details, which can be mainly extracted from the user's registration form, which exists in many e-commerce platforms:
\begin{enumerate}[resume]
	\item \textbf{Birth-Year($v$)} - the declared birth year of $v$.
	\item \textbf{State($v$)} - the declared state of $v$. For consistency, we converted the feature State($v$) to an integer using CRC32 hash function~\cite{black1994fast}. 
	\item \textbf{Active-Days($v$)} - the number of days $v$ was active in the e-commerce platform. In this study, we calculate this feature by calculating the number of days that passed between the date the user created a profile in the e-commerce platform and the last date the user performed a selling or buying transaction.
	
\end{enumerate}

\textbf{Target class.} 
Every instance in the training set includes a binary target attribute that indicates whether the user was identified as a shill bidder. 
In this study, we assumed that a list of identified shill bidders (referred to as \textit{shill bidders list}) is provided. 
These shill bidders are identified by the e-commerce platform experts and are used to train the supervised learning algorithms. 
A part of this list is also used as a ground truth during evaluation of the algorithms as described below.

Although, there are unidentified shill bidders that do not appear in the shill bidders list, as we see next, the fraction such users is expected to be extremely low. 
Therefore, for the purpose of training supervised learning algorithms we assume that a randomly picked user is a benign user if it does not appear in the shill bidders list.

\subsection{Selecting Users for the Training Set}
In this study we assumed that the actual ratio between shill bidders and benign users in an e-commerce platform is unknown, yet there are indications that the ratio between benign users and shill bidders is much in favor of the benign users. 
For example, in other online platforms, such as online social networks, there is a clear indication that the ratio between the number of benign entities and the number of malicious entities is in favor of the benign entities. 
The official Facebook estimation is that approximately 3-4\% of Facebook users are malicious users~\cite{Facebook18:online}, and according to Rahman et al.~\cite{rahman2012frappe} about 13\% of applications in Facebook are malicious. 
To construct our classifiers, we choose to create a balanced training set with an equal number of benign users and shill bidders, similar to the methodology used by Guha et al.~\cite{guha2004propagation} to predict trust, by Leskovec et al.~\cite{leskovec2010predicting} to predict positive and negative links, and by Fire et al.~\cite{fire2013friend} to identify fake users in Facebook.

\subsection{Choosing a Supervised Learning Algorithm}
\label{sec:alg}
To identify the supervised learning algorithm which yields the best classification results on our datasets, we trained the classifiers on all the users in the provided shill bidders list and an equal number of benign users. We then extracted for each user all the 31 features, which were described in Section~\ref{sec:features}. 
Afterwards, we used the constructed balanced training set and fed it to Weka~\cite{weka}, a popular suite of machine learning software.
We used Weka's OneR, C4.5 (J48) decision tree, K-Nearest-Neighbors (IBk; with K=3), Naive-Bayes,
Random-Forest, LogitBoost, Rotation-Forest, and Bagging implementations of the corresponding algorithms. For each of these algorithms, all of the configurable parameters were set to their default values.

We evaluated each classifier using the 10-fold cross validation method and calculated
the True-Positive (TP) rate, False-Positive (FP) rate, F-Measure (FM) value, and the Area-Under-Curve (AUC) measure.
These metrics assisted us in selecting the best supervised learning algorithm for identifying shill bidders.


\subsection{Supervised Learning Algorithm Evaluation}
\label{sec:ml_eval}
To evaluate the precision of the selected supervised learning algorithm, which received the highest AUC in ``the wild'' on a real e-commerce imbalanced dataset, we performed the following steps:\footnote{An additional method for evaluating the classifier precision is to manually validate the classifiers results. However, due to privacy limitations, these types of methods were not available in this study. }

\begin{enumerate}
	
	\item We created a balanced training set by randomly selecting 90\% of the shill bidder users in the shill bidders list, and by randomly selecting an equal number of benign users.

	\item The remaining 10\% of the users in the shill bidders list, which were excluded in step 1, were utilized to construct several test sets having various imbalance rates. 
	We created five test sets with 2, 5, 10, 20, and 100 benign users per single shill bidder. 
	
	\item For each user in the constructed training and testing sets, we extracted all the features, which were described in Section~\ref{sec:features}.
	
	\item Using the balanced training set and the selected supervised learning algorithm, we constructed a shill bidder identification classifier. 
	
	\item For each user in each one of the five imbalanced testing sets, we used the constructed balanced classifier to predict the user likelihood of being a shill bidder.
	
	\item For each imbalanced testing set, we calculated the classifier precision at the top \textit{k} (\textit{precision@k}). Namely, for each one of the five imbalanced datasets and for an integer $k \in [1,n]$, we calculated the percent of the top \textit{k} users which received the highest likelihood of being shill bidders and were actually shill bidders.
	
	\item Lastly, to reduce variability, we repeated steps 1 to 6 three times, and averaged the obtained precision at $k$ results for each $k$.
\end{enumerate}

\section{Empirical Evaluation}
\label{sec:empirical}
\subsection{The E-commerce Dataset}
\label{sec:dataset}
In this study, we used anonymized datasets provided by one of the largest e-commerce companies in the world to evaluate our shill-bidder detection algorithms.
Additionally, in order to query the provided datasets, we used Hadoop infrastructure which includes several thousands of Hadoop nodes.
The e-commerce dataset included information of several billions of buying and selling transactions, as well as several billions feedback transactions.
Each feedback transaction included feedback ratings with one of three possible values: negative feedback (-1), neutral feedback (0), and positive feedback (+1).
All of the transactions in the e-commerce dataset were actual transactions performed by over several hundred million platform users through the end of 2012. 
Furthermore, we were provided with a list of 187,224 user which were marked as shill bidders by the company's proprietary algorithms.
To select a benign users list for our training and testing sets throughout this study, we randomly selected from the e-commerce dataset a list of 500,000 seller-users which performed at least one sell transaction. 
We then removed from the benign users list all the users which also appeared in the shill bidders list.

\subsection{Experiment Setup}
\label{sec:setup}
We evaluated various supervised learning algorithms to construct classifiers which can identify which of the users are shill bidders.
Using the provided 187,224 shill bidders and an additional 187,224 benign users from the e-commerce dataset, we constructed a balanced training set and evaluated various supervised learning algorithms (see Section~\ref{sec:alg}).
Furthermore, using the e-commerce dataset, we constructed five imbalanced testing sets with the following ratios: 
\begin{enumerate}
\item \textit{1 to 2} - with 18,722 shill bidders and 37,444 benign users.
\item \textit{1 to 5} - with 18,722 shill bidders and 93,610 benign users.
\item \textit{1 to 10} - with 18,722 shill bidders and 187,220 benign users.
\item \textit{1 to 20} - with 15,000 shill bidders and 300,000 benign users.
\item \textit{1 to 100} - with 3,200 shill bidders and 320,000 benign users.
\end{enumerate}
 We used these imbalanced datasets to evaluate the constructed classifiers' precision at $k$ for $k \in [1, 30, 1000]$, using the methods described in Section~\ref{sec:ml_eval}.

\subsection{Results}
\label{sec:results}
In this section, we present the results obtained by the method described in Section~\ref{sec:methods}. 

\begin{figure*}[]
	
	\begin{center}
		\includegraphics[width=\textwidth]{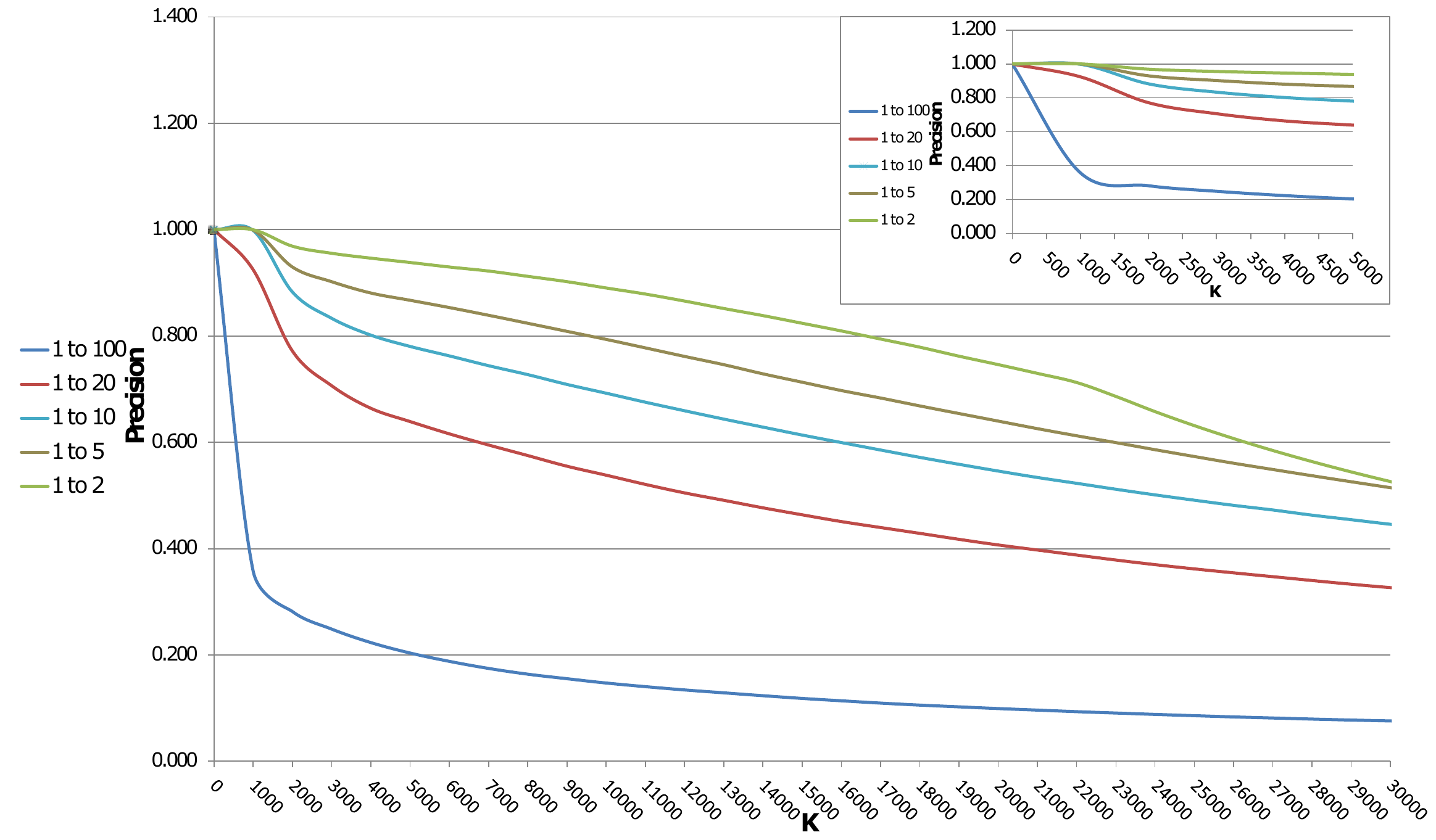}
	\end{center}
	\caption{ \label{fig:p_at_k} Rotation-Forest classifier precision at $k$ results - for five imbalance testing sets.
	}
\end{figure*}

\begin{table}
	\centering
	\caption{Supervised Learning Classifers Results on Balanced Training Set \label{tab:ml_balance}}
	\begin{tabular}{lllll}
		\hline 
		
		\textbf{Classifier}         & \textbf{TP}    & \textbf{FP}    & \textbf{FN}    & \textbf{AUC}   \\ \rowcolor{LightCyan}
		\hline	
		\textbf{OneR}               & 0.800 & 0.252 & 0.780 & 0.774 \\
		\textbf{Naïve-Bayes}        & 0.886 & 0.764 & 0.645 & 0.752 \\ \rowcolor{LightCyan}
		\textbf{Decision-Tree(J48)} & 0.822 & 0.193 & 0.816 & 0.860 \\
		\textbf{Random-Forest}      & 0.854 & 0.230 & 0.820 & 0.885 \\ \rowcolor{LightCyan}
		\textbf{Bagging}            & 0.834 & 0.179 & 0.829 & 0.902 \\
		\textbf{LogitBoost}         & 0.811 & 0.170 & 0.819 & 0.901 \\ \rowcolor{LightCyan}
		\textbf{Rotation-Forest}    & 0.845 & 0.173 & 0.838 & 0.912 \\
		\hline
		
	\end{tabular}
\end{table}
According to our evaluation results, among all tested supervised learning algorithms, the Rotation-Forest classifiers performed
best on the e-commerce dataset, with an especially good AUC result of 0.912 and a TP rate of 0.845 (see Table~\ref{tab:ml_balance}). 
Therefore, we chose to construct shill the bidder identification classifiers using the Rotation-Forest algorithm and to evaluate the constructed classifiers' precision at $k$ for the five imbalanced testing sets, which were defined in Section~\ref{sec:ml_eval}.
The results showing the classifiers' precision at $k$ average values for all five testing sets are presented in Figure~\ref{fig:p_at_k}.
According to the evaluation results, the developed shill bidders identification classifiers presented high precision at $k$, with precision at 1,000 of 1, 1, 0.999, 0.925, and 0.358, when the ratio between the number of shill bidder and the number of benign users was 1 to 2, 1 to 5, 1 to 10, 1 to 20, and 1 to 100, respectively (see Figure~\ref{fig:p_at_k}).
These results indicate that the presented shill bidder identification algorithms can detect shill bidders far better than a random algorithm. Moreover, the presented algorithm gave very high precision rates when the percentage of the shill bidder in the e-commerce dataset was at least 5\%.

\begin{table}[!htb]
    \begin{minipage}{.5\linewidth}
      	\caption{Features' Information Gain Scores \label{tab:infogain}}
      \tiny
      \centering
	\begin{tabular}{p{2.5cm} p{1.8cm}}
		\hline
		          & \textbf{Info Gain Score} \\
		\hline \rowcolor{LightCyan}
		\textbf{Min-Sell-Price}      & 0.268           \\
		\textbf{Sell-Trans-Num}      & 0.248           \\ \rowcolor{LightCyan}
		\textbf{Total-Sell-Quantity} & 0.247           \\
		\textbf{Unique-Buyers}       & 0.240           \\ \rowcolor{LightCyan}
		\textbf{State}               & 0.240           \\
		\textbf{Total-Sell-Amount }  & 0.196           \\ \rowcolor{LightCyan}
		\textbf{Rcv-Unique-Fdbk }    & 0.188           \\
		\textbf{Rcv-Fdbk-Num}        & 0.187           \\ \rowcolor{LightCyan}
		\textbf{Min-Buy-Price}       & 0.180           \\
		\textbf{Rcv-Fdbk-RSum}       & 0.162           \\ \rowcolor{LightCyan}
		\textbf{Rcv-Pos-Fdbk}        & 0.162           \\
		\textbf{Gvn-Unique-Fdbk}     & 0.154           \\ \rowcolor{LightCyan}
		\textbf{Gvn-Fdbk-Num}        & 0.153           \\
		\textbf{Gvn-Pos-Fdbk }       & 0.153           \\ \rowcolor{LightCyan}
		\textbf{Gvn-Fdbk-RSum}       & 0.153           \\
		\textbf{Fdbk-Bi-Degree}      & 0.150           \\ \rowcolor{LightCyan}
		\textbf{Total-Buy-Quantity } & 0.142           \\
		\textbf{Buy-Trans-Num}       & 0.143           \\ \rowcolor{LightCyan}
		\textbf{Unique-Sellers}      & 0.128           \\
		\textbf{Max-Sell-Price}      & 0.116           \\ \rowcolor{LightCyan}
		\textbf{Total-Buy-Amount}    & 0.120           \\
		\textbf{Rcv-Fdbk-Avg}        & 0.112           \\ \rowcolor{LightCyan}
		\textbf{Gvn-Fdbk-Avg}        & 0.089           \\
		\textbf{Max-Buy-Price}       & 0.076           \\ \rowcolor{LightCyan}
		\textbf{Active-Days}         & 0.079           \\
		\textbf{Max-Buy-Quantity}    & 0.070           \\ \rowcolor{LightCyan}
		\textbf{Gvn-Neg-Fdbk}        & 0.059           \\
		\textbf{Rcv-Neg-Fdbk}        & 0.047           \\ \rowcolor{LightCyan}
		\textbf{Max-Sell-Quantity}   & 0.016           \\
		\textbf{Birth-Year}          & 0.006           \\ \rowcolor{LightCyan}
		\textbf{Trans-Bi-Degree}     & 0.002          \\ 
		\hline
	\end{tabular}
    \end{minipage}%
    \begin{minipage}{.5\linewidth}
      \centering
      \tiny
        \caption{Features Average and Median Values Ratio Between Shill Bidders and Random Seller-Users  \label{tab:features}}
\begin{tabular}{lll}
		\hline
		& \textbf{AverageRatio} & \textbf{Median  Ratio} \\
		\hline \rowcolor{LightCyan}
		\textbf{Buy-Trans-Num}       & 2.689        & 4.421         \\
		\textbf{Sell-Trans-Number}   & 3.341        & 16.75         \\ \rowcolor{LightCyan}
		\textbf{Unique-Sellers}      & 2.642        & 4.094         \\
		\textbf{Unique-Buyers}       & 3.506        & 13.625        \\ \rowcolor{LightCyan}
		\textbf{Bidir-Trans-Users}   & 1.412        & 2             \\
		\textbf{Max-Buy-Price}       & 1.663        & 2.168         \\ \rowcolor{LightCyan}
		\textbf{Min-Buy-Price}       & 0.035        & 0.966         \\
		\textbf{Max-Buy-Quantity}    & 1.866        & 1.5           \\ \rowcolor{LightCyan}
		\textbf{Total-Buy-Quantity}  & 2.683        & 4.525         \\
		\textbf{Total-Buy-Amount}    & 2.441        & 4.295         \\ \rowcolor{LightCyan}
		\textbf{Max-Sell-Price}      & 2.693        & 2.676         \\
		\textbf{Min-Sell-Price}      & 0.063        & 0.227         \\ \rowcolor{LightCyan}
		\textbf{Max-Sell-Quantity}   & 1.373        & 1             \\
		\textbf{Total-Sell-Quantity} & 3.36         & 16.875        \\ \rowcolor{LightCyan}
		\textbf{Total-Sell-Amount}   & 3.583        & 9.874         \\
		\textbf{Gvn-Fdbk-Num}        & 2.76         & 6.286         \\ \rowcolor{LightCyan}
		\textbf{Rcv-Fdbk-Num}        & 2.964        & 6.463         \\
		\textbf{Gvn-Unique-Fdbk}     & 2.921        & 6.054         \\ \rowcolor{LightCyan}
		\textbf{Rcv-Unique-Fdbk}     & 3.112        & 6.222         \\
		\textbf{Bidir-Fdbk-Users}    & 2.989        & 6.091         \\ \rowcolor{LightCyan}
		\textbf{Gvn-Pos-Fdbk}        & 2.76         & 6.366         \\
		\textbf{Gvn-Neg-Fdbk}        & 2.748        & inf           \\ \rowcolor{LightCyan}
		\textbf{Rcv-Pos-Fdbk}        & 2.786        & 5.702         \\
		\textbf{Rcv-Neg-Fdbk}        & 2.04         & inf           \\ \rowcolor{LightCyan}
		\textbf{Gvn-Fdbk-RSum}       & 2.76         & 6.317         \\
		\textbf{Rcv-Fdbk-RSum}       & 2.789        & 5.66          \\ \rowcolor{LightCyan}
		\textbf{Gvn-Fdbk-Avg}        & 1.017        & 0.991         \\
		\textbf{Rcv-Fdbk-Avg}        & 1.025        & 0.994         \\ \rowcolor{LightCyan}
		\textbf{Birth-Year}          & 1            & 0.999         \\
		\textbf{Active-Days}         & 1.387        & 1.545         \\ \rowcolor{LightCyan}
		\textbf{Birth-Year}          & 1            & 0.999         \\
		\textbf{Active-Days}         & 1.387        & 1.545        \\ \rowcolor{LightCyan}
		\hline
	\end{tabular}
    \end{minipage} 
\end{table}

\section{The Shill Bidder Ecosystem}
\label{sec:eco_results}
\label{sec:eco}
In this study, we utilize the provided shill bidders list to empirically analyze the characteristics of shill bidders and to study the shill bidder ecosystem, i.e., the interactions of shill bidders with each other.

To analyze the shill bidders' characteristics using the e-commerce dataset, we calculated the average and median values for all the numeric features we defined in Section~\ref{sec:features} for the users in the provided shill bidders list. Additionally, to better understand which features assist the supervised learning algorithms to distinguish between shill bidders and benign, we used the balanced training set (see Section~\ref{sec:setup}) to calculate the features importance using Weka's Information Gain features' selection algorithm.
The results of the shill bidders' characteristics analysis are presented in 
Tables~\ref{tab:infogain} and~\ref{tab:features}.

From the shill bidders characteristics analysis results presented in Table~\ref{tab:features}, it can be observed that shill bidders behaved differently from the random seller-users in the following manner: First, on average, shill bidders are active for more days and perform far more selling and buying transactions than random sellers.
These results may indicate that shill bidder users are, in general, active users, which perform many buying and selling transactions.
Second, on average, shill bidders sell more products to unique buyers and buy more products from unique sellers than random seller-users.
However, the shill bidders' buying and selling minimum price was on average much less than the buying and selling minimum price of random-seller users.
We believe that these results are due to the shill bidders' attempts to maximum their profits and minimize their losses when the buy or sell feedback.
Third, on average, shill bidders received more negative feedback than random seller-users.
We assume that these results indicate that in the end many shill bidders utilize their obtained positive feedback to mislead other users in the platform, which in return gives the shill bidders negative feedback. 
Lastly, on average, shill bidders gave more negative feedback than random seller-users.
We believe that this result may indicate that shill bidders are also being utilized to perform sybil attacks~\cite{levine2006survey}.
We hope to prove these assumptions in a future study.
Additionally, we discovered that the shill bidders had 798 unique State feature values, while the randomly selected seller-users had 1,435 unique State feature values.
Furthermore, the most common State feature value among the shill bidders was the ``default'' value which appeared in the details of 134,979 shill bidders, while ``default'' state value appeared only in the details of 46,805 randomly selected seller-users.

From the features' Information Gain scores results, which are presented in Table~\ref{tab:infogain}, it can be observed that the Min-Sell-Price, Sell-Trans-Num, and Total-Sell-Quantity features received the highest Information Gain scores. We believe that these features received the highest scores due to the shill bidders behavioral patterns. In many cases, shill bidders attempted to decrease their losses by selling cheap products using many transactions in order to collect a great deal of positive feedback, to spend as little money as possible. 

To study the shill bidder ecosystem, we analyzed the feedback graph created by the shill bidders.
The shill bidder feedback graph can assist us in understanding the ``big picture'' beyond the shill bidders and their interactions, and even assist us in understanding the shill bidders' working methods. 
We defined the feedback graph as following: Given a list $\hat{V} \subseteq V$ of e-commerce users, we can define $\hat{V}$'s feedback graph to be a weighted directed graph, where each directed weighted link is defined to be the amount of feedback user $u \in \hat{V}$ gave user $v \in \hat{V}$. Formally,
the feedback graph is defined to be $ H_{\hat{V}} := <\hat{V},E_{\hat{V}}>$, where each link $e_{\hat{V}} \in E_{\hat{V}}$ in the graph defined as
\[e_{\hat{V}}:= \{(u,v,w)| \exists u,v \in {\hat{V}} \mbox { and } w = |\{(u,v,r,d) \in E_F\}| \}.\] 
Using the provided shill bidders list $B$, we first constructed the feedback graph $H_{B} := <B,E_B>$ as explained above.
We then calculated various graph properties using graph theory algorithms. 
Namely, we mostly used the igraph software package~\cite{csardi2006igraph} to calculate the following graph's properties: 
\begin{enumerate}
	\item number of vertices.
	\item number of links.
	\item maximum and minimum link weight.
	\item average link weight.
	\item number of bidirectional links ($|\{(u,v,w) \in E_B| \exists (v,u,w) \in E_B\}|$ ).
	\item density.
	\item components number.
	\item largest component size.
\end{enumerate}
Additionally, we used the igraph implementation of the Bron-Kerbosch algorithm~\cite{bron1973algorithm} to find the maximal cliques in the graph\footnote{The used igraph cliques detections algorithm implementation treats the directed graph as an undirected graph.} and calculate their distribution. By identifying cliques, we can identify a group of shill bidders who worked together. Lastly, we used a random sample to select a list with an equal number of seller-users, and we compared the properties of the feedback graph created by the random sampled seller-users of those the feedback graph created by the shill bidders.

From the ecosystem analysis results, it can observed that the shill bidder feedback graph is a relatively dense graph that spawn 1,805,199 directed links between 156,769 shill bidders, with an average of 1.31 feedback occurrences per link (see Table~\ref{tab:graph} and Figure~\ref{fig:graph}).
Additionally, 79.09\% of the shill bidders are located in a single component. 
Furthermore, according the maximal clique detection results, it can be noticed that in contrast to the random user feedback graph, the shill bidder feedback graph consists of many cliques (see Table~\ref{tab:clique}).
These results may indicate that many shill bidders work together and assist each other to receive positive feedback.
Another alternative explanation is that shill bidders open several accounts and use them to boost their own reputation, or they sell feedback to other users. 
\begin{table}[ht]
	\centering
	\caption{Feedback Graphs Properties \label{tab:graph}}
	
	\begin{tabular}{lll}
		\hline \rowcolor{White}
		& \textbf{Shill Bidders} & \textbf{Random Users}  \\ 
		& \textbf{Feedback Graph} & \textbf{Feedback Graph} \\ 
		\hline \rowcolor{LightCyan}
		\textbf{Number of Users}                                                                       & 187,224                       & 187,224                       \\
		\begin{tabular}[c]{@{}l@{}}\textbf{Number of Feedback } \\ \textbf{Between Users}\end{tabular}         & 2,391,312                     & 91,863                        \\ \rowcolor{LightCyan}
		\begin{tabular}[c]{@{}l@{}}\textbf{Number of Positive}\\ \textbf{Feedback Between Users}\end{tabular}  & 2,373,993                     & 91,097                        \\
		\begin{tabular}[c]{@{}l@{}}\textbf{Number of Negative}\\ \textbf{Feedback  Between Users}\end{tabular} & 8,786                         & 341                           \\ \rowcolor{LightCyan}
		\textbf{Number of  Non-Isolated Users}                                                         & 156,769                       & 35,599                        \\
		\textbf{Number of Links}                                                                       & 1,805,199                     & 67,383                        \\ \rowcolor{LightCyan}
		\textbf{Average Link Weight}                                                                   & 1.31                          & 1.35                          \\
		\textbf{Max Link Weight}                                                                       & 914                           & 219                           \\ \rowcolor{LightCyan}
		\textbf{Min Link Weight}                                                                       & -11                           & -8                            \\
		\textbf{Number of  Bidirectional Links}                                                        & 1,675,411                     & 59,692                        \\ \rowcolor{LightCyan}
		\textbf{Density}                                                                               & $7.35\cdot 10^{-5}$   & $5.32\cdot10^{-5}$   \\
		\textbf{Component Number}                                                                      & 8,123                         & 9,309                         \\ \rowcolor{LightCyan}
		\textbf{Largest Component Size}                                                                & 148,072 (79.09\%)             & 21,443 (11.45\%)              \\
		\textbf{Maximal Clique Size}                                                                   & 7                             & 3                             \\ \rowcolor{LightCyan}
		\textbf{Maximal Clique Number}                                                                 & 895,844                       & 37,080\\
		\hline                       
	\end{tabular} 
\end{table}

\begin{table}[ht]
	\centering
	\caption{Maximal Cliques Size Distributions \label{tab:clique}}
	\begin{tabular}{lll}
		\hline
		\textbf{Clique} & \textbf{Random} & \textbf{Shill}  \\
		\textbf{Size} & \textbf{Feedback Graph} & \textbf{Feedback Graph}  \\
		\hline \rowcolor{LightCyan}
		\textbf{3} & 260    & 66,892 \\
		\textbf{4} & 0      & 3,945  \\ \rowcolor{LightCyan}
		\textbf{5} & 0      & 803    \\
		\textbf{6} & 0      & 173    \\ \rowcolor{LightCyan}
		\textbf{7} & 0      & 23    \\
		\hline
	\end{tabular}
\end{table}

\section{Conclusions}
\label{sec:conclusion}
According to the presented method various user features are first  extracted from user transactions (buying and selling) and from feedback activities (giving and receiving). 
Second, supervised learning algorithms are used to train a model for classifying users into shill bidders and legitimate accounts.
We evaluated the algorithms using a real large-scale anonymized e-commerce dataset. 
This dataset includes over several billions of buying and selling occurrences performed by several hundred million users, as well as several billions of feedback occurrences they gave or received through the end of 2012. 
The dataset also includes a list of 187,224 users, which were marked as shill bidders in e-commerce platform systems and were used as ground truth for training and testing our algorithms. 
Evaluation results of the presented method show the area under the ROC curve (AUC) of up to 0.912 and precision at 1,000 of 0.999 when the ratio between the shill bidders and the benign users was 1 to 10 (see Section~\ref{sec:results}).

By analyzing the e-commerce dataset, we also empirically studied the characteristics of shill bidders, as well as the shill bidder ecosystem.
As a result of this analysis, we discovered that, on average, shill bidders were more active, performed more selling and buying of transactions, and gave and received more feedback compared to randomly selected seller-users (see Table~\ref{tab:features}).
Moreover, we also discovered that shill bidders gave more negative feedback compared to randomly selected seller-users.
These results may indicate that shill bidders can be used to also perform sybil attacks~\cite{levine2006survey}.
These sybil attacks can be against targeted platform users, such as shill bidders competitors, and aim to damage the targeted users' reputations by giving them unjust negative feedback.
Additionally, we discovered that the shill bidder feedback graph, which was constructed from all the feedback links between each two shill bidders (see Section~\ref{sec:eco}), is a relatively dense graph with 1,805,199 links among 156,769 shill bidders (see Table~\ref{tab:graph} and Figure~\ref{fig:graph}).
Furthermore, in the shill bidders feedback graph we identified 66,892 cliques with at least 3 shill bidders, and 23 cliques with at least 7 shill bidders (see Table~\ref{tab:clique}).
These results indicate that many shill users collaborate to increase their overall reputations. Moreover, these results may indicate the existence of e-commerce bots which perform automatic buying and selling transactions.
These bots also submit feedback to each other and to other users. 

We believe that these observations regarding the shill bidder ecosystem can assist in improving the detection of shill bidders in following manner: First, we can utilize the results regarding the shill bidder graph structure and improve our shill bidder identification classifiers by extracting additional graph structure-based features, such as the number of cliques a user is member of and the number of shill bidders the user is connected to.
Second, we can use the fact that many shill bidders are connected to each other (see Figure~\ref{fig:graph} and Table~\ref{tab:graph}) and set our shill bidder identification classifier to identify shill bidders among users which are connected to several shill bidders, instead of applying the classifier on random sets of users chosen out of several hundred million e-commerce platform users.
We believe that this technique of focusing the shill bidder identification classifier on these connected users can identify shill bidders with even higher precision.
Lastly, we can utilize the results that indicate the shill bidders tend to formulate relatively large cliques (see Table~\ref{tab:clique}) and use various clique identification algorithms to identify large cliques in the feedback graph created from all the e-commerce users' feedback interactions.
We believe that large cliques in this feedback graph have a high likelihood of containing shill bidders. 
We hope to verify these three assumptions in our future research.

The study presented here offers many additional future research directions to pursue.
One possible research direction is to analyze not only the structured data which was extracted from the e-commerce users' feedback and transactions, but also to use Natural Language Processing (NLP) algorithms to analyze the users' content data, such as the feedback comments and the products information page. 
An additional possible research direction is to utilize various clustering algorithms to identify shill bidders.
Another possible research direction is to construct classifiers, which utilize similar features to those presented in Section~\ref{sec:features}, to identify other types of malicious users, such as fraudsters who sell fictional products. 

\bibliographystyle{splncs04}
\bibliography{usenix}

\end{document}